\documentclass[aps,pra,twocolumn,amsmath,amssymb,showpacs,superscriptaddress]{revtex4}

\newcommand{\bra}[1]{\langle#1|}
\newcommand{\ket}[1]{|#1\rangle}

\usepackage[dvips]{graphicx}
\usepackage{mathrsfs}

\begin{document}

\bibliographystyle{apsrev}

\title{Strategies for the preparation of large cluster states using non-deterministic gates}

\author{Peter P. Rohde}
\email[]{rohde@physics.uq.edu.au}
\homepage{http://www.physics.uq.edu.au/people/rohde/}
\affiliation{Centre for Quantum Computer Technology, Department of Physics\\ University of Queensland, Brisbane, QLD 4072, Australia}

\author{Sean D. Barrett}
\affiliation{Blackett Laboratory, Imperial College London, Prince Consort Road, London SW7 2BW, United Kingdom}

\date{\today}

\frenchspacing

\begin{abstract}
The cluster state model for quantum computation has paved the way for schemes that allow scalable quantum computing, even when using non-deterministic quantum gates. Here the initial step is to prepare a large entangled state using non-deterministic gates. A key question in this context is the relative efficiencies of different `strategies', i.e. in what order should the non-deterministic gates be applied, in order to maximize the size of the resulting cluster states? In this paper we consider this issue in the context of `large' cluster states. Specifically, we assume an unlimited resource of qubits and ask what the steady state rate at which `large' clusters are prepared from this resource is, given an entangling gate with particular characteristics. We measure this rate in terms of the number of entangling gate operations that are applied. Our approach works for a variety of different entangling gate types, with arbitrary failure probability. Our results indicate that strategies whereby one preferentially bonds together identical qubits are considerably more efficient than those in which one does not. Additionally, compared to earlier analytic results, our numerical study offers substantially improved resource scaling.
\end{abstract}


\maketitle

\section{Introduction}
The development of cluster state (or one-way) quantum computation \cite{bib:Raussendorf01} has presented us with a completely alternate perspective to the standard circuit model for quantum computing (QC). This model has the benefit that it transfers all entangling quantum gates to an offline state preparation stage. This has attracted much interest in the quantum computing community because in some instances it has highly favourable properties. One of the key benefits offered by this model is its applicability to architectures where entangling gates are non-deterministic. Here the cluster state model can allow significant physical resource savings when compared to the circuit model, most notably in optical implementations \cite{bib:Nielsen04, bib:YoranReznik03, bib:BarrettKok05, bib:LimBarrett05, bib:Duan05, bib:BrowneRudolph05}. The central idea here is that a large cluster state, a highly entangled state of many qubits, can be `grown' using a `divide-and-conquer' approach, whereby smaller clusters are iteratively bonded together, a process which can lead to very efficient physical resource scaling characteristics. While several approaches for efficiently growing cluster states using non-deterministic gates have been prescribed, a key open question is ``what are the minimal physical resource requirements for preparing a certain cluster using entangling gates with a given success probability?'' and ``what strategy (or algorithm) must be employed to achieve optimal resource scaling?''.

A number of other authors have considered the problem of efficiently constructing cluster states using non-deterministic two-qubit entangling operations. Refs.~\cite{bib:YoranReznik03, bib:Nielsen04} present procedures for generating multi-qubit entangled states (cluster states in the case of Ref.~\cite{bib:Nielsen04} and `linked states' in Ref.~\cite{bib:YoranReznik03}) using non-deterministic linear optical gates, and calculate the efficiency of this process for particular values of the success probability of the elementary entangling operations, $p_\mathrm{gate}$, with $p_\mathrm{gate} > 1/2$. Subsequently, in Ref.~\cite{bib:BrowneRudolph05}, a simplified linear optical scheme for generating cluster states was described, and it was also shown how to scalably construct clusters in the case where $p_\mathrm{gate} = 1/2$. In Ref.~\cite{bib:RalphHayes05} a comparable scheme is described within the standard circuit model.

In Ref.~\cite{bib:BarrettKok05}, and subsequently Ref.~\cite{bib:Duan05}, these ideas where extended to the case of arbitrary values of $p_\mathrm{gate}$. In particular, it was shown that, even for arbitrarily small values of $p_\mathrm{gate}$, linear clusters can be grown with a cost that grows linearly with the size of the required cluster, albeit with a linear coefficient that becomes rather large for very small values of $p_\mathrm{gate}$.

One shortcoming of the calculations presented in Refs.~\cite{bib:BarrettKok05,bib:Duan05} is that, in order to obtain analytic approximations for this overhead cost, certain assumptions are made that necessarily reduce the efficiency of the protocol. In particular, it is assumed that short clusters (below some critical length that depends on $p_\mathrm{gate}$) are constructed without recycling -- upon gate failure, the remaining sub-clusters are discarded and construction starts from scratch. In this paper, we avoid such profligate measures by directly computing the cost of a more frugal scheme via numerical simulations.

In general, the questions of optimal strategies and physical resource requirements are difficult problems. Previous work by Kieling \emph{et al.} \cite{bib:Kieling06,bib:Gross06} considered this issue. Here the authors specifically focussed on the linear optics implementation of Ref.~\cite{bib:BrowneRudolph05}, where gate success probability is assumed to be $p_\mathrm{gate}=1/2$. They assumed a resource of $N$ Bell pairs, and asked, for a given strategy of bonding them, what the expected length of the longest chain, $Q(N)$, will be. They then considered the behavior of $Q(N)$ against $N$ for various strategies and asked what the upper bound on this scaling relationship was. There are two strategies that were given special attention in their analysis that we will repeat here. The first is {\sc Greed}, whereby we always preferentially bond the largest existing cluster states together. Intuitively one might expect such a strategy to perform extremely well -- we rapidly build up clusters of higher length. The second is {\sc Modesty}, where we do exactly the opposite and always preferentially bond the smallest available clusters together. Surprisingly, the authors showed that {\sc Greed} is in fact not only a sub-optimal strategy, but {\sc Modesty} is substantially better and close to optimal. More recently, other strategies for growing cluster states based on percolation theory have been examined \cite{bib:KielingRudolphEisert06}.

In this paper we consider the question of cluster state preparation strategies from a slightly different perspective. Specifically, if we are ever to implement large scale cluster state quantum computing we will effectively be operating in a regime where we must have an effectively infinite resource of qubits. In such a situation we are interested in the rate at which we can prepare `large' clusters from this resource, i.e. how many elementary entangling operations must we perform per large cluster we prepare? Thus, in our analysis we assume an infinite resource of qubits and calculate the rate at which large clusters are prepared from this resource, given a particular entangling gate. Our approach is to employ a numerical Monte-Carlo simulation of the preparation procedure. This technique works for a variety of different entangling gate types with arbitrary gate success probability.

In this paper we focus on the preparation of linear clusters, which are basic primitive in the construction of larger clusters. There are many strategies for preparing arbitrary large cluster states. For example, in the scheme of Nielsen \cite{bib:Nielsen04}, a resource of star-shaped `micro-cluster' is used to prepare large cluster states using non-deterministic gates. These star-shaped clusters can in turn be prepared from linear clusters using local operations and single qubit measurements. Similarly, in the scheme of Duan \& Raussendorf \cite{bib:Duan05} a resource of linear clusters are used to prepare `+'-shaped clusters, which facilitate the construction of lattice clusters using non-deterministic gates. The scheme of Kieling et al. \cite{bib:Kieling06,bib:Gross06} describes a `weaving' technique, where a lattice cluster is prepared directly from a number of linear clusters. A different approach to preparing lattice clusters from linear clusters is described in Ref.~\cite{bib:BrowneRudolph05,bib:BarrettKok05,bib:LimBarrett05,bib:Benjamin05}. In these cases the additional cost of preparing 2D clusters from 1D chains is well understood. Therefore, it is appropriate to focus our attention on the preparation of linear clusters as a primitive for use in such higher-level protocols.

This paper is structured as follows. In Section~\ref{sec:background} we present background material on cluster states, strategies and gate types. In Section~\ref{sec:analysis} we describe our analysis techniques, and in Section~\ref{sec:results} we discuss our simulation results. We conclude in Section~\ref{sec:conclusion}.

\section{Background} \label{sec:background}

\subsection{Cluster states}
The standard circuit model for quantum computation is very analogous to our usual understanding of classical circuits -- we begin with an input state, apply a series of gates to manipulate the state, and then measure the output to determine the `answer'. The cluster state model is completely different and has no analogy in classical circuit theory. Here we prepare a highly entangled state, known as a cluster (or graph) state, and then perform some sequence of single qubit measurements. The choice of measurements -- both the measurement basis and the order in which they are performed -- then implements the algorithm.

It has been shown that lattice clusters are universal for quantum computations up to a given size. Thus lattice clusters act as a resource for universal quantum computation. For this reason theoretical studies often focus on lattice clusters.

\begin{figure}[!htb]
\includegraphics[scale=0.45]{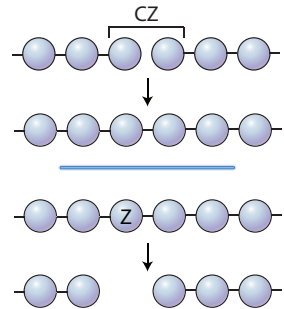}
\caption{Creating links in a cluster state using a CZ gate (top). Removing a qubit from a cluster state with a $Z$ measurement (bottom).} \label{fig:cluster_manipulations}
\end{figure}

A cluster state can be expressed as a graph, hence the alternate term `graph state'. In this representation vertices represent single qubits that are initially prepared in the $\ket{+}=(\ket{0}+\ket{1})/\sqrt{2}$ state. Graph edges represent controlled-phase (CZ) gates performed between the neighbouring vertices. This bonding procedure is shown in Fig.~\ref{fig:cluster_manipulations}.

An additional property of cluster states that we will make use of in this paper is qubit removal. When a qubit is measured in the $Z$ eigenbasis this removes the qubit and any edges connecting to it, while leaving the remainder of the cluster unchanged. This property is useful to recover from certain types of gate failure. For example, consider a linear cluster to which we attempt to bond another cluster at the end. If the bonding fails with no well-defined failure outcome (i.e. the qubit is effectively depolarized) then performing $Z$ measurements on the qubits adjacent to the affected ones recovers the remainder of the cluster by isolating the malignant qubits from the rest of the graph. This is shown in Fig.~\ref{fig:cluster_manipulations}.

We have only recapped the basic properties of cluster states requisite for our analysis techniques. Excellent reviews of the cluster state model can be found in Refs.~\cite{bib:Raussendorf01, bib:Raussendorf03, bib:Nielsen06}

\subsection{Strategies}
Beginning with a resource of qubits, we begin bonding cluster together to construct progressively larger cluster states. At any point in time we will have a `pool' of clusters of varying lengths, to which we may apply bonding operations. A `strategy' is an algorithm that inspects the current pool of available clusters and decides which two clusters to attempt to bond together. Previously we reviewed two strategies -- {\sc Greed} and {\sc Modesty}. In our simulations we will consider several additional strategies, which we summarize in Table~\ref{table:strategies} (Note that the {\sc Modesty} strategy applied in this work differs from that in Ref. \cite{bib:Kieling06,bib:Gross06}, as will be discussed in Section \ref{sec:results}).
\begin{table}
\begin{tabular}{|c|c|}
\hline
Strategy	         & Choice of clusters to bond \\
\hline
{\sc Greed}          & Two largest available clusters\\
{\sc Modesty}        & Two smallest available clusters\\
{\sc Random}         & Randomly choose two clusters\\
{\sc Paired Greed}   & Largest available identical clusters\\
{\sc Paired Modesty} & Smallest available identical clusters\\
{\sc Paired Random}  & Random identical clusters\\
\hline
\end{tabular}
\caption{Summary of different bonding strategies. For the {\sc Modesty} and {\sc Paired Modesty} strategies single qubits are excluded from the decision making process, unless no larger clusters are available. This is necessary since we assume an infinite resource of single qubits. This is distinct from the {\sc Modesty} strategy employed in Ref. \cite{bib:Kieling06,bib:Gross06}. The {\sc Paired} variations of strategies always only bond two identical clusters, and bins containing only a single chain will undergo bonding.} \label{table:strategies}
\end{table}

Typically there will be much room for parallelizability in bonding strategies, limited only by the number of gates available, the number of clusters in the pool, and potentially other practical constraints. While this parallelisation will not reduce the overall number of bonding operations that must be performed, it will, in general, reduce the average amount of time qubits spend in quantum memory. Clearly this will generally be highly favorable from a decoherence perspective.

In all our simulations do not employ any parallelization. That is, exactly one gate operation is applied per time step. There are two reasons for this. First, it eliminates a parameter and makes the results much easier to digest. Second, the resource efficiency of a strategy without parallelization will represent an upper bound on the achievable resource efficiency with parallelization, and therefore is an extremely useful parameter to know. This is because any strategy involving parallelization can be simulated without parallelization, but the converse is not true in general. Therefore, parallelizability represents an added constraint on the class of strategies that can effectively be implemented.

\subsection{Gate types} \label{sec:gate_types}
\begin{figure}[!htb]
\includegraphics[scale=0.45]{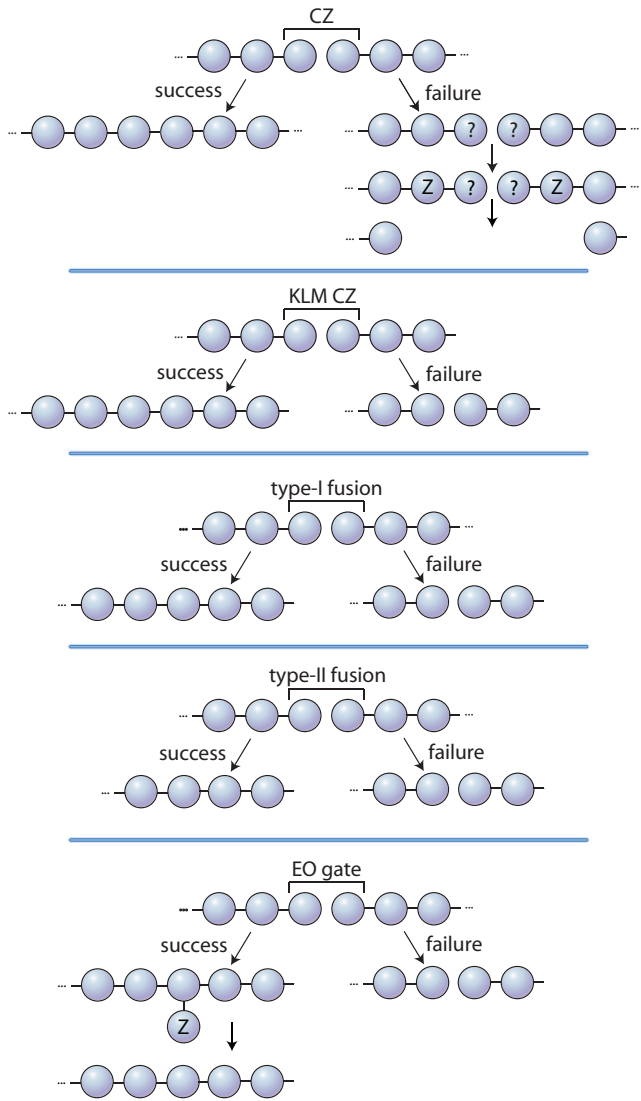}
\caption{Bonding of linear clusters using the CZ, KLM CZ, type-I fusion, type-II fusion, and EO gates.} \label{fig:gates}
\end{figure}

In addition to various strategies for bonding clusters together there are a variety of gates that may be employed to perform the necessary bonding operations to prepare cluster states. Specifically, there are two features that differentiate bonding gates. The first is their success probability, $p_\mathrm{gate}$. In our simulations we treat this as a parameter. The second is their resource usage. That is, how qubits are wasted upon success and failure respectively. Formally, a bonding operation will act on a set of two linear clusters of lengths $l_1$ and $l_2$, which, upon success or failure will be mapped to a set of new linear clusters. In Table~\ref{table:gates} we summarize the most well known gate types and their resource usage. The action of these gates is shown graphically in Fig.~\ref{fig:gates}.

In our simulations we will primarily focus on three gates in particular: two variations of the CZ gate -- the archetypical bonding gate; and, the so-called EO (Entangling Operation) gate introduced in Ref.~\cite{bib:BarrettKok05}. These gates are particularly relevant to current architectures for performing quantum computation using non-deterministic gates.

The regular CZ gate, upon success, acts in the usual way, implementing a non-destructive controlled-phase operation. Upon failure its action is undefined. Thus to recover the remaining cluster one must measure the qubits neighboring the ones acted upon by the CZ gate in the $Z$-basis. This disconnects those qubits from the cluster resulting in two clusters, each two qubits smaller in size \cite{bib:Duan06}.

The KLM CZ gate differs in that upon failure it performs $Z$-measurements on the affected qubits. Thus, these qubits are automatically removed from the clusters and no additional measurements are needed to recover. Therefore, the KLM CZ gate is more resource savvy and, upon failure, results in two clusters each with only one qubit removed.

The EO gate is employed in a hybrid scheme, involving both photons and `matter qubit' systems. These may be trapped ions or atomic systems, impurities in semiconductors or insulators (e.g the NV-diamond system) or quantum dots. In this approach, the matter systems contain additional optical transitions, such that photons can be emitted whose state is conditional on the logical qubit state. In Ref.~\cite{bib:BarrettKok05} the photons are encoded in separate time windows, although other encodings such as frequency have subsequently been proposed \cite{bib:Duan06}.

Two matter qubit systems can be entangled by first inducing them to emit photons, passing these photons through a simple passive linear optical device (typically a beam splitter), and detecting the output modes with destructive photo-detectors. In Ref.~\cite{bib:BarrettKok05} two such rounds are implemented, one for each time window. Success of the scheme is `heralded' by a particular sequence of detector clicks, and results in a highly entangled state of the two qubits.

In fact, it can be shown that the scheme amounts to a partial parity measurement. That is, in the case of success, a projection operator $\Pi_+ = \ket{01}\bra{01} + \ket{10}\bra{10}$ (up to local unitary corrections) is applied to the state of the two
qubits. In the case of failure, each of the two qubits may be affected by a Pauli $Z$ (phase flip) error. Success occurs with
probability $p_\mathrm{gate} = \eta^2/2$, where $\eta$ is the combined photon collection and detection efficiency for each of the modes output from the beam splitter.

This partial parity measurement can be used to generate linear cluster states as follows. Pairs of unentangled qubits are first
prepared in the product state $\ket{+}\ket{+}$. Applying the EO, followed by single qubit local unitary corrections \cite{bib:BarrettKok05} leads to a two qubit cluster state in the case of success. In the case of failure, the matter qubits are in an unknown state and should be re-initialised. Given an $l_1$-qubit linear cluster state, a single qubit can be added to one end of the chain by preparing the qubit in the state $|+\rangle$ and applying the EO between this qubit and the qubit at the end of the chain. On success, this results (up to local unitary corrections) in a chain of length $l_1+1$. On failure, the qubit at the end of the linear cluster may have undergone a $Z$-error, and so this qubit should be removed from the cluster by measuring it in the $Z$-basis. The result is a chain of length $l_1-1$ and two unentangled qubits which may be re-prepared. Finally, in this paper we will often consider the case when two chains of arbitrary lengths, $l_1$ and $l_2$, are fused at their endpoints. In this case, on success, the result (up to local unitary corrections) is a chain with a `backbone' of length $l_1+l_2-1$ with a further dangling node (or `cherry') which is an extra qubit connected to the backbone at the point where the two chains were connected \cite{bib:Benjamin05}. For the purposes of this work, we assume for simplicity of our calculations that this cherry is removed via a $Z$-measurement. However, in general the cherry may be utilized to increase the efficiency of growing linear clusters or higher dimensional graph states \cite{bib:Benjamin05}.
Thus the estimates of the cost presented in this paper may be reduced even further.

On failure, again the end qubits of each chain must be measured in the $Z$-basis, and the result is two chains of length $l_1-1$ and $l_2-1$, and two unentangled qubits. These results are summarized in Table~\ref{table:gates}. Note that, apart from the initial step of making two qubit clusters, the resource usage rules for the EO and the type-I fusion \cite{bib:BrowneRudolph05} are identical.

\begin{figure}
\begin{tabular}{|c|c|c|}
\hline
Gate type                 & Success         & Failure \\
\hline
CZ                        & $\{l_1+l_2\}$   & $\{l_1-2,l_2-2\}$ \\
KLM CZ                    & $\{l_1+l_2\}$   & $\{l_1-1,l_2-1\}$ \\
Type-I fusion             & $\{l_1+l_2-1\}$ & $\{l_1-1,l_2-1\}$ \\
EO (for $l_2=1$)          & $\{l_1+1\}$     & $\{l_1-1,1,1\}$   \\
EO (for $l_2>1$)          & $\{l_1+l_2-1\}$ & $\{l_1-1,l_2-1\}$ \\
Type-II fusion            & $\{l_1+l_2-2\}$ & $\{l_1-1,l_2-1\}$ \\
\hline
\end{tabular}
\caption{Resource usage of different cluster state bonding operations, acting on the initial cluster states $\{l_1,l_2\}$. `CZ' is a generic non-destructive controlled-sign gate. No qubits are lost upon success. However, the failure mode of the gate is undefined. Thus, recovering the clusters upon failure requires measuring the neighbouring qubits in the $Z$ eigenbasis \cite{bib:Duan05}. The `KLM CZ' gate is a Knill-Laflamme-Milburn \cite{bib:KLM01} type controlled-phase gate which performs $Z$ measurements on both qubits upon failure. This has the effect of removing the qubits from the cluster and no further recovery measurements are necessary. The `type-I fusion' \cite{bib:BrowneRudolph05} and `EO' \cite{bib:BarrettKok05} gates implement a partially destructive fusion. Upon success one of the two qubits is consumed, and upon failure both qubits are effectively measured in the $Z$ eigenbasis. The `type-II fusion' gate differs only in that upon success both qubits are consumed.} \label{table:gates}
\end{figure}

\section{Analysis} \label{sec:analysis}
We now turn our attention to our analysis technique. Our analysis proceeds as follows. We begin with a pool of linear clusters of various lengths. We describe the state of the pool by a population vector $\vec{n}$, whose elements correspond to the number of clusters in the pool of the respective length. Beginning with an initial population vector $\vec{n}_0$, we repeatedly apply a bonding strategy, which inspects the population vector and decides which two chains to attempt to bond together using the given gate operation. Let $f_s$ represent one iteration of the strategy. Then the evolution of the pool of clusters takes the form,
\begin{equation}
\vec{n}_{t+1} = f_s(\vec{n}_t).
\end{equation}
As discussed previously, we always assume strategies apply exactly one gate operation per time unit. Thus, $t$ gives the total number of gate operations that are applied.

In our simulations we assume the population vector is initialized such that all elements are initially zero, except the first element (i.e. $l=1$, the single qubit bin), which is set to infinity. As we begin applying a given strategy to the initial population vector, higher length bins begin to populate, and eventually longer chains will begin `spilling over' off the end of the vector. We regard any of these spillover chains as being `complete'. Note that the definition of `complete' depends on how many bins are used in the simulation, which we label $L$. We tally the total number of qubits that spill out of the population vector, which we label $N_Q$.

After a sufficient number of applications of the bonding strategy the system reaches a quasi steady-state, whereby the average population of each bin in the population vector remains constant, averaged over a sufficiently large time period \footnote{Within the average timescale it takes a qubit to exit the system, the dynamics are not steady state. However, averaged over a much larger timescale than this the discrete dynamics of individual qubits hopping between bins averages out.}. In our simulations we are interested in the quasi steady-state dynamics of this evolution. In particular we are interested in the rate at which large clusters are prepared per gate operation. Thus we define,
\begin{equation} \label{eq:rate_def}
r = \lim_{t\to\infty}\frac{N_Q(t)}{t}.
\end{equation}
Note that $1/r$ corresponds to the average number of attempted entangling gate operation per qubit added to the completed chains. At first our treatment of large clusters may appear somewhat dubious, in that we are treating all large clusters, irrespective or their individual lengths, as being equal and simply putting them in one basket. The key observation here is that the resource overhead in bonding together multiple large clusters is negligible compared to the size of the cluster. Specifically, with a gate success probability of $p_\mathrm{gate}$ one expects to perform $1/p_\mathrm{gate}$ bonding operations on average until two clusters are successfully joined. If the lengths of the individual chains are much larger than this, $l_{1,2}\gg 1/p_\mathrm{gate}$, this overhead is negligible and bonding can essentially be treated as `free'. Thus, in the large cluster regime, it is justified to assume that a pool of large clusters is equivalent to a single large cluster of size equal to the cumulative size of the pool. Furthermore, for $L\gg 1/p_\mathrm{gate}$, $r$ becomes independent of $L$. Thus, for simulation purposes we can simply choose some value of $L$, sufficiently large that changes in $L$ leave $r$ unchanged.

It is important to note that this technique is only suited to the analysis of linear cluster state preparation. The reason for this is that the population vector, $\vec{n}$, contains a single entry for each possible graph configuration. When restricted to linear graphs the dimensionality of $\vec{n}$ is just the maximum number of qubits. However, in the 2D case the dimensionality of $\vec{n}$ grows exponentially with the number of qubits, ruling out even moderately large simulations.

\section{Results} \label{sec:results}
We simulate the effect of bonding strategies and gate parameters using a Monte-Carlo approach. We use a population vector with $L=50$ bins, and simulate over 50,000 time steps. At each time step we tally $N_Q$, the number of spillover qubits. Finally, at the end of the simulation we calculate $r$, the rate at which large clusters were prepared. We consider all six strategies listed in Table~\ref{table:strategies}. There is one caveat associated with the {\sc Modesty} and {\sc Paired Modesty} strategies. Specifically, we exclude the single qubit bin from the calculation, unless no other bins are populated. The reason for this variation, relative to the {\sc Modesty} strategy discussed in Ref. \cite{bib:Kieling06,bib:Gross06}, is that we are assuming an infinite resource of single qubits. Thus, without this modification {\sc Modesty} strategies would forever be stuck bonding single qubits together.

\begin{figure}[!htb]
\includegraphics[width=\columnwidth]{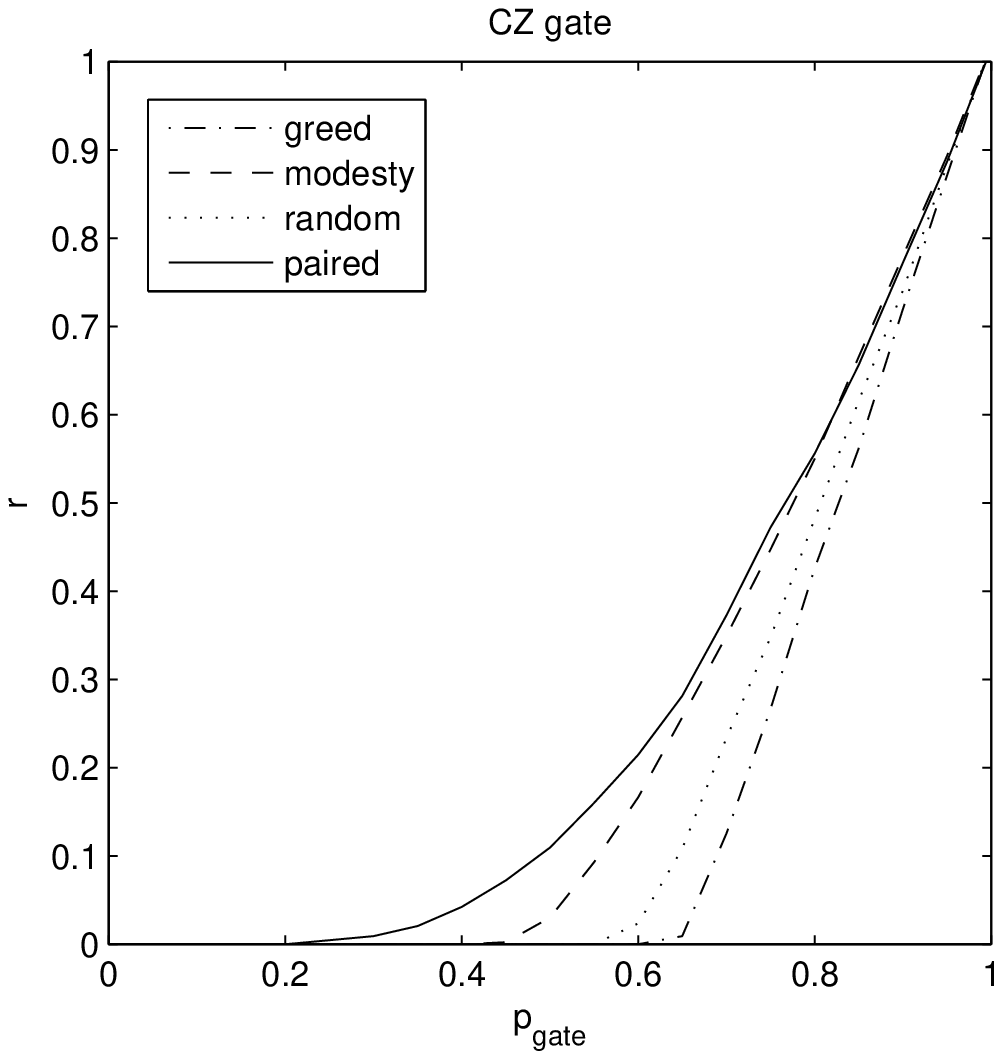}\\
\includegraphics[width=\columnwidth]{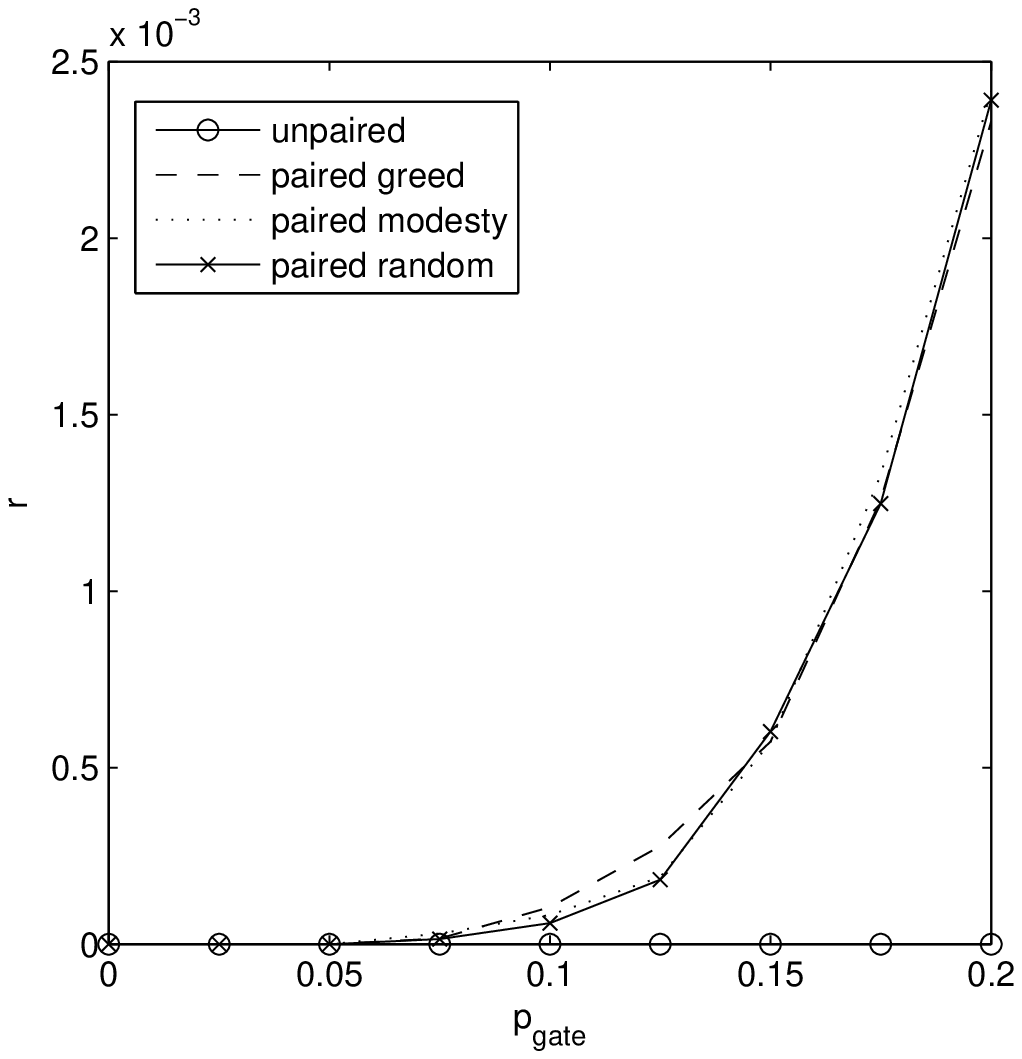}
\caption{Spillover rate against gate success probability for all six bonding strategies. The population vector has size $L=50$, and we apply a Monte-Carlo simulation over 50,000 time steps ($5\times 10^6$ for lower plot). The lower plot is zoomed in on the region $p_\mathrm{gate}<0.2$. All three {\sc Paired} strategies exhibit identical performance, which has been labeled {\sc Paired}.} \label{fig:spillover_vs_p}
\end{figure}

In Fig.~\ref{fig:spillover_vs_p} we plot $r$ against $p_\mathrm{gate}$ for all strategies.
There are several interesting features. In agreement with \cite{bib:Kieling06,bib:Gross06} the {\sc Modesty} strategy performs better than {\sc Greed} for all $p_\mathrm{gate}$. The {\sc Random} strategy performs worse again. Particularly interesting are the three {\sc Paired} strategies, which all exhibit identical performance within the accuracy of our simulation. Perhaps surprisingly the {\sc Paired} strategies perform significantly better than the un-{\sc Paired} ones. In particular, for small values of $p_\mathrm{gate}$, $r$ drops to zero (within the accuracy our simulations) for all the un-{\sc Paired} strategies, which is not the case for the {\sc Paired} strategies.

\begin{figure}[!htb]
\includegraphics[width=\columnwidth]{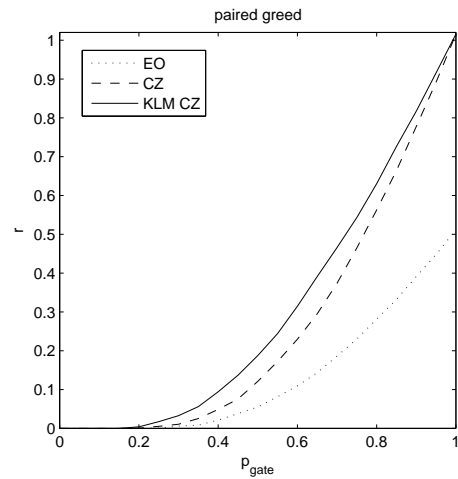}
\caption{Spillover rate against gate success probability for three different gate types -- CZ, KLM CZ and EO. All plots are using the {\sc Paired Greed} strategy. The population vector has size $L=40$, and we apply a Monte-Carlo simulation over 100,000 time steps. Note that in practise the EO gate has a maximum success probability of $p=1/2$.} \label{fig:gate_types}
\end{figure}

We have compared the performance of a variety of strategies for a given gate, the CZ gate. Next we consider the performance of different gates for a fixed strategy. In Fig.~\ref{fig:gate_types} we plot the performance of the {\sc Paired Greed} strategy for three different gate types: CZ, KLM CZ and EO.

Consider the general behavior of these plots. In the limit $p_\mathrm{gate}=1$ the rate of large cluster preparation is $r=1$. This is expected. It tells us that once the system attains its quasi-steady state, large clusters are spilling off the end of the population vector at an average rate of one qubit per gate operation, which clearly must necessarily be the case. For $p_\mathrm{gate}<1$ we observe a lower $r$. This is because now the gate has some failure probability which results in larger clusters being converted into smaller cluster with non-zero probability. In other words, the dynamics of the population vector are no longer described by a one-way coupling from lower bins to higher bins, but now includes reverse couplings which slow down the overall flux of qubits towards higher length.

The most striking feature of these plots is that all three {\sc Paired} strategies exhibit identical performance, which is significantly better than any of the other un-{\sc Paired} strategies. Although we considered only a limited set of strategies, empirically this suggests that one should always preferentially bond clusters of equal length, and beyond this the performance is strategy independent.


\subsection{Comparison with analytic results}
We finish by comparing the scaling relationships obtained through our numerical methods with two recent analytical studies of resource scaling. In the first instance we compare against the analytic results of Duan \& Raussendorf \cite{bib:Duan05} for preparing linear clusters using non-deterministic CZ gates. In the second instance we compare against the analytic expressions obtained by Barrett \& Kok for preparing linear clusters using the EO gate. See Appendix \ref{app:analytic_estimates} for a discussion on the derivation of the analytic expressions.
\begin{figure}
\includegraphics[width=\columnwidth]{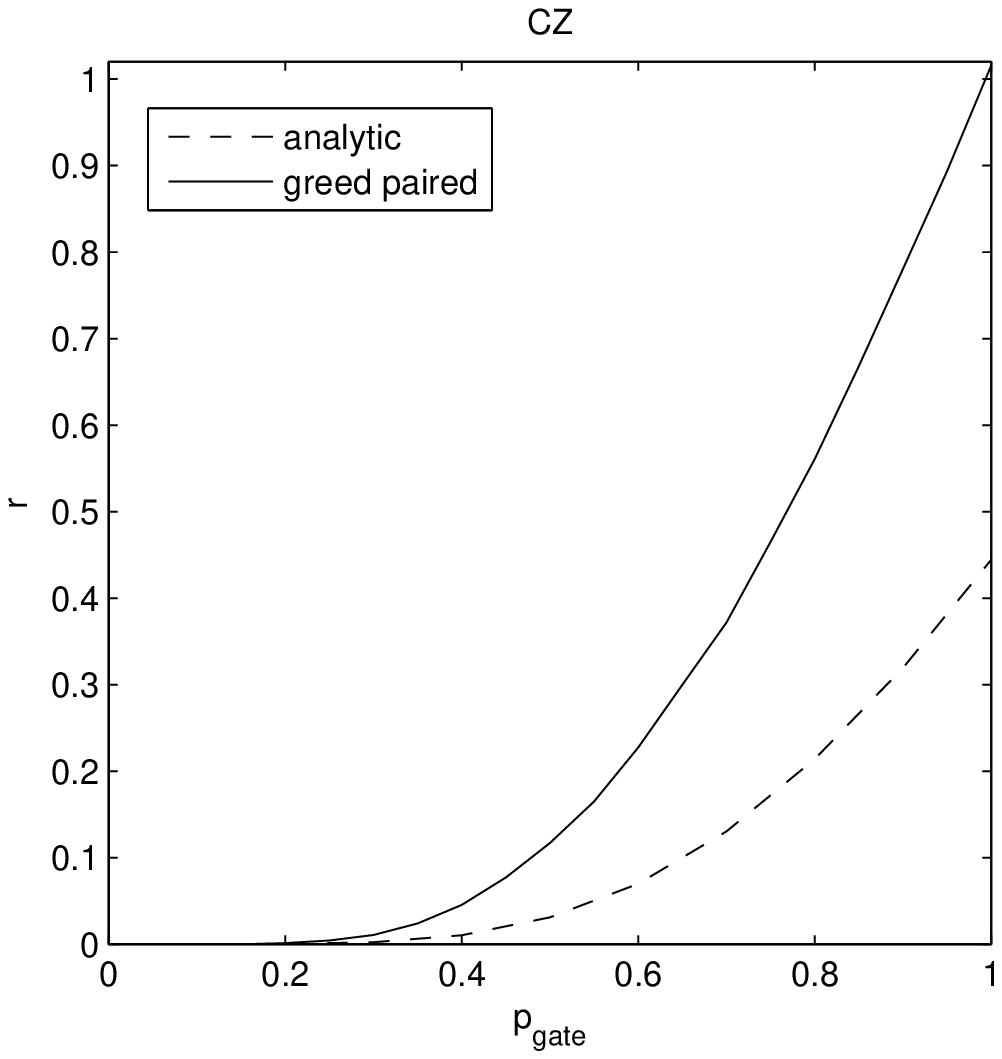}
\includegraphics[width=\columnwidth]{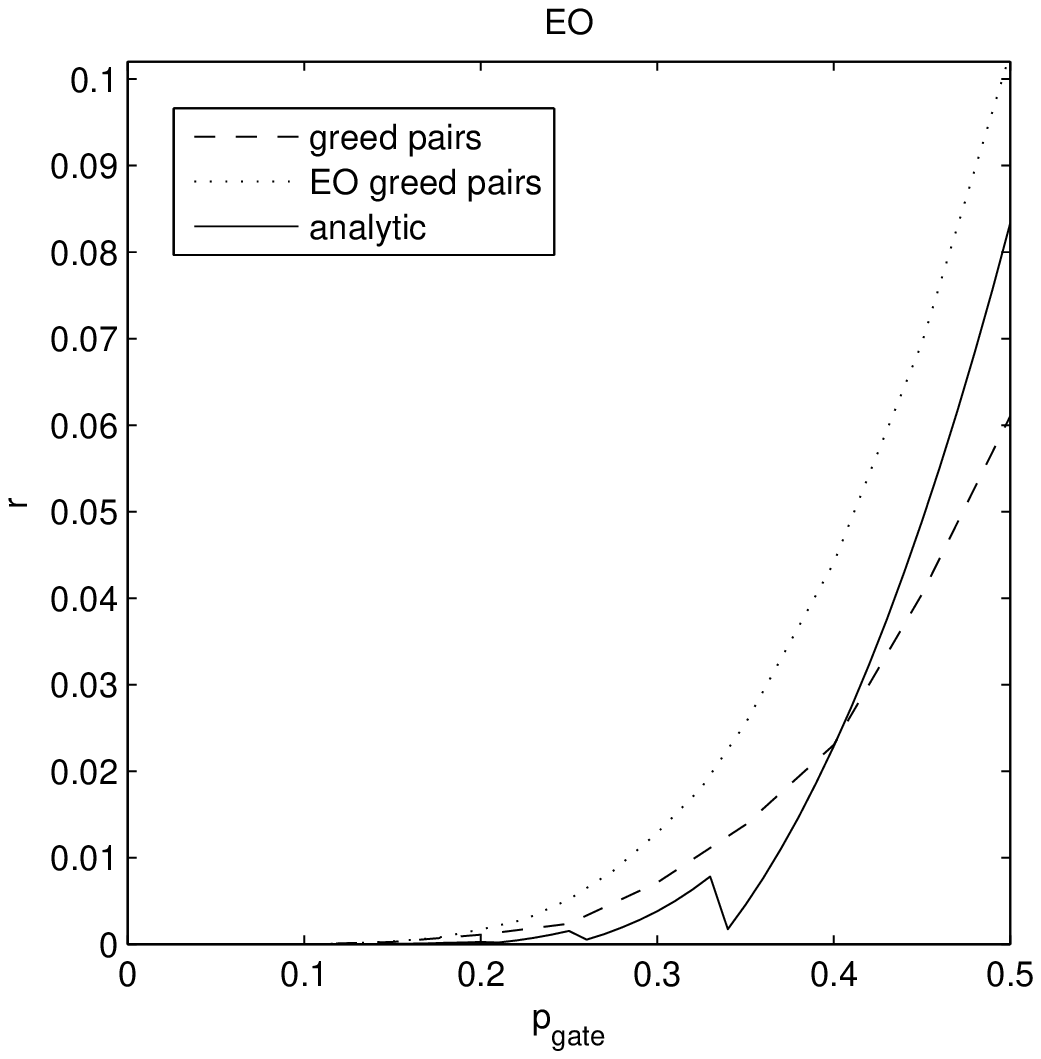}
\caption{(top) Numerical scaling relationship for the CZ gate compared with the analytic scaling relationship obtained by Duan \& Raussendorf. (bottom) Two numerical scaling relationships for the {\sc EO} gate compared with the analytic scaling relationship obtained by Barrett \& Kok. The fist numerical simulation is for the normal {\sc Greed Paired} strategy. The second, {\sc EO Greed Paired}, is a modification of this where we always preferentially bond $\{l_1=2,l_2=1\}$ pairs.} \label{fig:analytic_vs_numerical}
\end{figure}
Note that in both plots the best numerical scaling relationships are significantly more efficient than the analytic ones, owing to the lack of recycling in the analytic derivations leading to inefficient resource usage. In the case of the EO gate we have included a third strategy, labeled {\sc EO Greed Paired}, which is identical to the usual {\sc Greed Paired} strategy except that if a Bell pair is available we always preferentially bond it with a single qubit to attempt to form a GHZ state. The reason for this exception is that with this type of gate bonding two Bell pairs together also forms a GHZ state upon success. Thus, doing so wastes the operation required to form one of the Bell states. At higher levels this discrepancy no longer occurs, so the usual pairing approach is favorable.

\section{Conclusion} \label{sec:conclusion}
We have considered the question of preparing large cluster states using non-deterministic and potentially destructive bonding operations, and the effect that bonding strategies have on the resource requirements of this process. Empirically, our results indicate that {\sc Paired} strategies, i.e. ones where we only bond clusters of equal length, perform much better than un-{\sc Paired} ones, especially in the regime of small $p_\mathrm{gate}$. Additionally, all three {\sc Paired} strategies considered exhibited identical performance. This suggests that the performance of large cluster state preparation is strategy independent provided one only bonds chains of equal size. This is in stark contrast to un-{\sc Paired} strategies, where performance is highly strategy dependent. In this case it is observed that the {\sc Modesty} strategy is superior to a {\sc Greed} approach, consistent with the observations made in Ref.~\cite{bib:Kieling06,bib:Gross06}.

We also compared the {\sc Paired} strategies with earlier analytic estimates for the cost of preparing linear clusters. We found substantial improvement in efficiency over these earlier estimates. We attribute this to the use of full recycling in our numerical analysis, whereas previous analytic results did not make full use of cluster recycling. This suggests that numerical techniques play an important role in the evaluation of resource scaling efficiency.

The analysis techniques we described are very general and hold for arbitrary gate success probabilities and a variety of different entangling gate types, including destructive ones. Therefore, our approach is suited to calculating physical resource requirements in a diverse array of different situations. However, our technique does have the disadvantage that it is only well suited to directly analyzing the preparation of linear clusters.

Importantly, in our analysis we have not \emph{proven} which strategies are optimal, we have simply made empirical observations that some classes of strategies are better than others, and that some are equivalent. Thus there is much room for further analysis in this direction. The MATLAB source code used for our simulations is available at {\tt http://www.physics.uq.edu.au/people/rohde/}.

\begin{acknowledgments}
We thank the members of the ``QiSci questions'' meeting at the University of Queensland, where aspects of these calculations were discussed. We also thank Pieter Kok and Jens Eisert for useful discussions on aspects of this work. PR acknowledges supported from the Australian Research Council, Queensland State Government, and DTO-funded U.S. Army Research Office Contract No. W911NF-05-0397. SDB is supported by the EPSRC.
\end{acknowledgments}

\appendix

\section{Analytic estimates of the cost of growing cluster states with non-deterministic operations} \label{app:analytic_estimates}

An analytic estimate of the cost of growing cluster states using non-deterministic operations was given in Ref. \cite{bib:BarrettKok05}. For the purposes of obtaining simple expressions for the cost of making linear clusters, the method used was intrinsically less efficient than the one used in our numerical simulations in this work. Here, we elaborate on the calculation presented in Ref. \cite{bib:BarrettKok05}, for the purposes of a comparison with our numerical results.

The method used was to build short cluster chains (above some critical length, as described below) using an inefficient technique, and then join these chains together to form a long chain. Assume that we already have a linear chain of length $N$, and would like to add a short chain of length $m$ to one end of this chain, using the EO with success probability $p_\mathrm{gate}$. Using the rules given in Table \ref{table:gates}, the expected length of the chain after this operation is
\begin{eqnarray}
L &=& p_\mathrm{gate} (N+m-1) + (1 - p_\mathrm{gate}) (N-1) \nonumber\\
&=& N+ m p_\mathrm{gate} - 1.
\end{eqnarray}
Thus, in order for the chain to grow on average, we require $L>N$ which implies that the short chains must satisfy
\begin{equation} \label{eq:crit_len_cond}
m > \frac{1}{p_{\textrm{gate}}}\,.
\end{equation}
Therefore the critical length depends directly on the gate success probability.

To make these short chains of length $m$, we use a `divide and conquer' technique. For ease of calculation, we assume that if any of the gates fail, any cluster chain fragments are simply discarded, and the process starts from scratch. For instance, to make a chain of length $m=3$ qubits, we attempt to bond 2-chains with single qubits. The total number of attempted gate operations to make a single 3-chain is then
\begin{eqnarray}
R_3 &=& \frac{ R_2 }{p_\mathrm{gate}}+ \frac{1}{p_\mathrm{gate}} \\
&=& \frac{1}{{p_\mathrm{gate}}^2} + \frac{1}{p_\mathrm{gate}} \,.
\end{eqnarray}
Here, $R_2 = 1/p_\mathrm{gate}$ is the number of attempted operations required to make a 2-chain. Similarly, we have for 4-chains
\begin{equation}
R_4 = \frac{1}{{p_\mathrm{gate}}^3}+\frac{1}{{p_\mathrm{gate}}^2}+ \frac{1}{p_\mathrm{gate}} \,.
\end{equation}

To make longer chains using this inefficient procedure, we can recursively join these shorter chains together. For instance, to make a 5-chain, two 3-chains can be joined together. Joining two 5-chains together results in a 9-chain, joining two 9-chains results in a 17-chain, and so on. If all this is done without recycling, the number of attempted operations to make 5-, 9-, and 17-chains, respectively, are given by
\begin{eqnarray}
R_5 &=& \frac{2}{{p_\mathrm{gate}}^3} + \frac{2}{{p_\mathrm{gate}}^2} + \frac{1}{p_\mathrm{gate}} \,,\nonumber\\
R_9 &=& \frac{4}{{p_\mathrm{gate}}^4} + \frac{4}{{p_\mathrm{gate}}^3} + \frac{2}{{p_\mathrm{gate}}^2} +
\frac{1}{p_\mathrm{gate}} \,, \nonumber\\
R_{17} &=& \frac{8}{{p_\mathrm{gate}}^5} + \frac{8}{{p_\mathrm{gate}}^4} + \frac{4}{{p_\mathrm{gate}}^3} + \frac{2}{{p_\mathrm{gate}}^2} + \frac{1}{p_\mathrm{gate}}.\nonumber\\
\end{eqnarray}
One can easily recurse these relations further to determine $R_m$ for larger values of $m$ (corresponding to smaller values of $p_\mathrm{gate}$).

Once we have a short chain of sufficient length (i.e. satisfying Eq. \ref{eq:crit_len_cond}, we can attempt to join the chain to the large chain of length $N$. The expected number of qubits added to the long chain is thus $m p_{\textrm{gate}}-1$, at a cost of $R_m + 1$ attempted operations. The total cost is therefore
\begin{equation}
C_m = \frac{R_m +1}{mp - 1}
\end{equation}
attempted operations per qubit added to the long chain. Note that the dominant term in the denominator of this expression has an exponent that grows roughly logarithmically with ${p_\mathrm{gate}}^{-1}$. Thus, asymptotically, the cost grows slightly faster than polynomially with ${p_\mathrm{gate}}^{-1}$.

The particular choice of $m$ to be used depends on the value of $p_{\textrm{gate}}$. The reciprocal of $C_m$ can be compared directly with the spillover rate in our numerical calculations, and therefore in Fig. \ref{fig:analytic_vs_numerical} we plot the functions ${C_{17}}^{-1}$, ${C_9}^{-1}$, ${C_5}^{-1}$, ${C_4}^{-1}$ and ${C_{3}}^{-1}$ for the regions $1/17 < p_\mathrm{gate} \le 1/9, 1/9 < p_\mathrm{gate} \le 1/4, 1/4 < p_\mathrm{gate} \le 1/3$ and $1/3 < p_\mathrm{gate} \le 1/2$, respectively.

A similar calculation, which uses slightly more recycling and is therefore somewhat more efficient, was subsequently performed in Ref. \cite{bib:Duan05}. Their calculations correspond to the non-deterministic CZ gate of Table \ref{table:gates}. The approximate cost that they obtain for long linear chains can be determined from Eq. \ref{eq:rate_def} of Ref. \cite{bib:Duan05} and is given by
\begin{equation}
C \approx \frac{1}{2} \left(\frac{2}{p} \right)^{\log_2(4/p+1)} \,.
\end{equation}
Again, $C^{-1}$ can be compared directly with the spillover rate for the non-deterministic CZ gate, and is plotted in Fig. \ref{fig:analytic_vs_numerical}.

\bibliography{paper}

\end{document}